\documentstyle[12pt,aps]{revtex}
\newcommand{\bold}[1]{\mbox{\boldmath $#1$}}
\tighten
\begin{document}
\draft
\title{\flushleft On linear momentum in quasistatic electromagnetic
systems\footnote
{This comment is written by V Hnizdo in his private capacity. No support or
endorsement by the Centers for Disease Control and Prevention is intended or
should be inferred.
}}

\author{\flushleft V Hnizdo}

\address{\flushleft
National Institute for Occupational Safety and Health,\\
1095 Willowdale Road, Morgantown, WV 26505, USA
\newline
\newline
E-mail: vbh5@cdc.gov}

\address{\flushleft\rm
{\bf Abstract}. 
The analysis of Aguirregabiria, Hern\'andez, and Rivas 
[2004 {\it Eur. J. Phys.} {\bf 25} 555--567] of the electromagnetic 
linear momentum in quasistatic systems omits the presence of an 
equal and  opposite hidden mechanical momentum in such systems. 
The existence of hidden
mechanical momentum ensures that the action-reaction law holds for the forces
between the charge- and current-carrying bodies of a quasistatic system, and 
thus the total mechanical momentum associated with the motion of 
such bodies in a closed quasistatic system is conserved.} 

\maketitle

\section*{}
\noindent                                                       
In a recent article \cite{AHR}, Aguirregabiria, Hern\'andez and Rivas (AHR)
consider examples of quasistatic systems consisting of current-carrying 
bodies and moving charges, and assert: 
`In all of them the internal electromagnetic forces will
not satisfy the action--reaction principle, so that to maintain constant the
total mechanical momentum one has to apply a non-null external force on the 
system. Then, the only way to keep the balance equation (the total external 
force equals the derivative of the total linear momentum) is to attribute a 
linear momentum to the [quasistatic] electromagnetic field.' AHR then
calculate the electromagnetic linear momentum in their examples,
and show that its time derivative  indeed equals the external force that they
believe is required in order to keep the mechanical momentum 
associated with the motion of the bodies in the system constant.

We would like to point out that such an explanation  of the necessity 
for a non-zero  momentum of a quasistatic electromagnetic field 
is misleading because it ignores the existence of an equal and
opposite mechanical momentum that is not associated with
any `overt' motion of the bodies in a quasistatic system. This quantity,
called hidden momentum,
was `discovered' some 35 years ago in investigations of the forces that
an electrically neutral current loop may experience in an electric field
\cite{Shockley,HP}. The existence and importance of hidden momentum 
in quasistatic electromagnetic systems are now universally accepted,
and this topic is included in the last editions of authoritative
textbooks \cite{Jackson,Griffiths}.
The surprising fact that a current-carrying body 
the centre of mass of which is at rest may possess a nonzero mechanical 
momentum is demanded by a general theorem that requires that the 
total (i.e., `electromagnetic' plus `mechanical') momentum of any finite
stationary distribution of charge, current and matter must vanish 
\cite{CV,VH3}; the hidden mechanical momentum required to match a nonzero 
momentum of a  quasistatic electromagnetic field can be shown to arise as a 
subtle relativistic effect from the electric-current-constituting motion of 
charged elements of the current-carrying body \cite{VH3,Vaid}.
(Some quantities, like for example the stress in a charged fluid medium, 
that may play an important role in the emergence of hidden momentum in
a given specific mechanism of the current transport are microscopically 
of an electromagnetic origin, and so the division of the physical 
quantities of any macroscopic system into `electromagnetic' and  `mechanical' 
is only conventional.) 

A quasistationary system with a current density $\bold{J}$ and 
electrostatic potential $\Phi$ contains a hidden momentum \cite{VH3}
\begin{equation}
\bold{P}_h=-\frac{1}{c^2}\int {\rm d}V\, \Phi\bold{J}
\label{Ph}
\end{equation}
which is equal and opposite to the momentum $\bold{P}_f$ of the quasistatic
fields $\bold{E}=-\bold{\nabla}\Phi$ and $\bold{B}$, 
$\bold{\nabla}\times\bold{B}=\mu_0\bold{J}$ in the system:
\begin{equation}
\bold{P}_f=\epsilon_0\int {\rm d}V\,\bold{E}\times\bold{B}
=\frac{1}{c^2}\int {\rm d}V\, \Phi\bold{J}.
\label{Pf}
\end{equation}
Here, the second equality can be derived using integration by parts
(see, e.g., \cite{VH3}). When the electric field varies only very 
little over the distribution of the current density $\bold{J}$,
the hidden momentum (\ref{Ph}), and thus also the negative of electromagnetic
momentum (\ref{Pf}), can be approximated by \cite{VH3,Vaid,VH0}
\begin{equation}
\bold{P}_h=\frac{1}{c^2}\,\bold{m}\times\bold{E}
\label{Pm}
\end{equation}
where $\bold{m}=\case{1}{2}\int {\rm d}V\,\bold{r}\times\bold{J}$ is the 
magnetic dipole moment of the electric currents and $\bold{E}$ is the value 
of the electric field at the currents. The presence of hidden momentum in a 
current-carrying body modifies the force that the body experiences
in external electric and magnetic fields. This is seen most easily by 
considering the force on a magnetic dipole moment $\bold{m}$ that is
located in external quasistatic magnetic ($\bold{B}$) and electric 
($\bold{E}$) fields, and whose velocity vanishes at the given instant of time.
The rate of change of the total (i.e., overt plus hidden) mechanical momentum 
of the dipole is given by the standard expression 
$\bold{\nabla}(\bold{m\cdot B})$, but that means that the force $\bold{F}$,
understood as the rate of change of the dipole's overt mechanical momentum,
i.e., that arising
from its translational motion, is diminished by the rate of change of  
the hidden momentum (\ref{Pm}) of the dipole \cite{HP,Vaid,VH0}:
\begin{equation}
\bold{F}=\bold{\nabla}(\bold{m\cdot B})
-\frac{1}{c^2}\,\frac{{\rm d}}{{\rm d}t}(\bold{m}\times\bold{E}).
\label{Fm}
\end{equation}
Using this expression for the force on a magnetic dipole, it is not 
difficult to show \cite{VH1} that the action--reaction principle holds 
true for the forces acting between the body that carries the currents
characterized by the magnetic moment $\bold{m}$ and a moving
charge creating the electric field $\bold{E}$.

The validity of the action--reaction principle for the interactions between
current-carrying bodies and charged bodies in a quasistationary system
is a general result \cite{Furry}, independent of the magnetic dipole
approximation used in (\ref{Fm}). The total momentum associated with 
the motion of such bodies in a closed quasistationary system is thus
conserved itself, and therefore the existence
of a linear momentum of a quasistatic electromagnetic field cannot
be detected through an overt-mechanical-momentum non-conservation 
\cite{VH1} (angular momentum of a quasistatic electromagnetic field
is different in this respect, as it can be detected through a
non-conservation of mechanical angular momentum---this is the famous
Feynman's disk `paradox' \cite{Feynman}---because there in no
equal and opposite hidden angular momentum to match the
electromagnetic angular momentum \cite{VH1}).

Taking the AHR's example of a thin toroidal solenoid and a point charge
$q$ moving along the toroid's axial symmetry axis, we can show 
easily that in fact the net force on the solenoid vanishes,
as required by the action--reaction principle and the fact that there is
no force on the charge since the toroidal solenoid does not produce any
magnetic field in its exterior. The hidden momentum $\bold{P}_h$ of the 
solenoid is 
evaluated most easily using expression (\ref{Pm}) in a line integral
along the torus main circuit $C$:
\begin{equation}
\bold{P}_h=\frac{1}{c^2}\oint_C {\rm d}\bold{m}\times\bold{E}
\label{Ps}
\end{equation}
where ${\rm d}\bold{m}=(NIS/2\pi){\rm d}\phi\,\hat{\bold{\phi}}$
is the magnetic dipole moment of an element of the solenoid 
subtended by an azimuth angle ${\rm d}\phi$ (cf.\ equation (14) of AHR), and
$\bold{E}{=}(1/4\pi\epsilon_0)[q/(R^2{+}a^2)]\hat{\bold{\zeta}}$
is the charge's electric field at ${\rm d}\bold{m}$ [$N$, $I$ and $S$ are
the solenoid's number of turns, current and cross-sectional
area, respectively, and $R$, $a$ and $\hat{\bold{\zeta}}$ are the toroid's
main radius, distance from its centre to the charge  and a unit vector
along the straight line passing through the charge and the element of dipole 
moment
${\rm d}\bold{m}$, respectively (see figure 1 of AHR)]. With the $z$-axis
along the straight line passing through the toroidal solenoid's centre and 
the charge,
only the $z$-component of (\ref{Ps}) is non-zero, giving
\begin{equation}
\bold{P}_h=-\frac{\mu_0}{4\pi}\,qNIS\,\frac{R}{(R^2+a^2)^{3/2}}\,\hat{\bold{z}}
\label{Pz}
\end{equation}
which is equal and opposite to the electromagnetic field momentum $\bold{P}_f$
of the system (see equation (21) of AHR). Taking into account the hidden
momentum (\ref{Pz}), the net force $\bold{F}$ on the solenoid is then
\begin{equation}
\bold{F}=\bold{F}_B-\frac{{\rm d}\bold{P}_h}{{\rm d}t}
=\bold{F}_B-\frac{3\mu_0}{4\pi}\,qNIS\,\frac{Ra\dot{a}}{(R^2+a^2)^{5/2}}
\,\hat{\bold{z}}
\label{Fs}
\end{equation}
where $\bold{F}_B$ is the direct force on the solenoid due to the magnetic 
field of the moving charge. But since the force $\bold{F}_B$ equals the second
term on the right-hand side of (\ref{Fs}) (see equation (18) of AHR),
the net force on the solenoid indeed vanishes,
\begin{equation}
\bold{F}=\bold{F}_B-\frac{{\rm d}\bold{P}_h}{{\rm d}t}=0.
\end{equation}

Contrary to an assertion of AHR, no external force is therefore necessary 
to be applied on the torodial solenoid in order to keep it at rest and thus
to maintain the total overt mechanical momentum of the system constant.
Hidden-momentum analysis of the other example of AHR will yield a similar 
conclusion that the effective force on the circuit is equal and opposite to
the magnetic force on the charge, and thus no external force is needed to be 
applied to any part of the system in order to maintain its
total overt mechanical momentum constant.

Before concluding, it should be mentioned  that a quasistationary system's 
hidden  momentum $\bold{P}_h$ and  electromagnetic  momentum $\bold{P}_f$ 
vanish when the body that supports the
current density $\bold{J}$ is a good electric conductor. Indeed, the
electrostatic potential $\Phi$ is constant in such a body and such a system,
and thus the integrals (\ref{Ph}) for $\bold{P}_h$ and (\ref{Pf}) for
$\bold{P}_f$ are  proportional to 
$\int{\rm d}V\bold{J}$, which vanishes as an integral over a divergenceless 
localized vector quantity. However, one can show that even then the force
on such a body is still given in the dipole approximation by equation 
(\ref{Fm}), where the second term on the right-hand side now arises from the
force on the  charge and/or current induced in the conductor by the external 
electric field in order to keep the potential $\Phi$ throughout the body 
of the conductor constant \cite{VH0}.

The momentum of the electromagnetic field of a quasistationary system does not
manifest itself through a violation of the conservation of overt mechanical 
momentum---its physical significance is rather more subtle as that of a quantity 
that is equal and opposite to the hidden mechanical momentum of the system.
Hidden momentum is an essential element in the momentum balance of a 
quasistationary electromagnetic system, and therefore its existence cannot be 
ignored in a correct momentum analysis of any such physical system, no matter 
how idealized or schematic it is for simplicity assumed to be. 
The existence of hidden momentum is necessary for maintaining the correct 
relativistic-transformation properties of the observed total energy, momentum
and rest mass of a charge- and current-carrying body \cite{VH2,VH4}, and, 
in fact, is not completely `hidden' since it manifests itself in the force 
that a current-carrying body experiences in an electromagnetic field.

\end{document}